# Fast unified reconstruction algorithm for conventional, phase-contrast and diffraction tomography

TIMUR E. GUREYEV,[1,2,*] , HAMISH G. BROWN[3], HARRY M. QUINEY,[1] AND LESLIE J. ALLEN[1]

[1]*School of Physics, University of Melbourne, Parkville, VIC 3010, Australia*
[2]*School of Physics and Astronomy, Monash University, Clayton, VIC 3800, Australia*
[3]*Ian Holmes Imaging Center, Bio21 Molecular Science and Biotechnology Institute, University of Melbourne, Parkville, VIC 3010, Australia*
*\*timur.gureyev@unimelb.edu.au*

**Abstract:** A unified method for three-dimensional reconstruction of objects from transmission images collected at multiple illumination directions is described. The method may be applicable to experimental conditions relevant to absorption-based, phase-contrast or diffraction imaging using X-rays, electrons and other forms of penetrating radiation or matter waves. Both the phase retrieval (also known as contrast transfer function correction) and the effect of Ewald sphere curvature (in the cases with a shallow depth of field and significant in-object diffraction) are incorporated in the proposed algorithm and can be taken into account. Multiple scattering is not treated explicitly, but can be mitigated as a result of angular averaging that constitutes an essential feature of the method. The corresponding numerical algorithm is based on three-dimensional gridding which allows for fast computational implementation, including a straightforward parallelization. The algorithm can be used with any scanning geometry involving plane-wave illumination. A software code implementing the proposed algorithm has been developed, tested on simulated and experimental image data and made publicly available.

## 1. Introduction

Methods for three-dimensional (3D) reconstruction of internal structure and composition of objects from transmission images collected at different illumination directions have been developed over the past 60 years or so, starting from the seminal work by Bracewell [1], Cormack [2], Crowther [3], Hounsfield [4] and others. These methods span a very broad range of experimental applications from radio astronomy to medical imaging and atomic-resolution transmission electron microscopy (TEM). A common mathematical formalism used in the majority of such methods is known as computed tomography (CT) [5], which can be traced back to the work by Radon published at the beginning of the twentieth century [6]. The original CT formalism corresponds to a physical imaging model where the radiation propagates through the imaged object along straight lines and the image contrast is the result of different attenuation of incident rays along different trajectories in accordance with the Beer-Lambert law [5]. A modified form of CT, called diffraction tomography (DT), was proposed by Emil Wolf [7] and later extended and refined by others [8-10]. Unlike the conventional CT, DT can explicitly take into account the Fresnel diffraction of the scattered waves inside the imaged object, which leads to deviation of the internal trajectories from straight lines. In the reciprocal (Fourier) space, this corresponds to incorporation of the Ewald sphere curvature into the treatment of the problem [11,12]. These effects become important in practice at high spatial resolutions, $\Delta$, and relatively long radiation wavelengths, $\lambda$, when the corresponding depth of field (DOF), $DOF = \Delta^2 / (2\lambda)$, becomes smaller than the extent of the object along the illumination direction [13,14] (Fig.1). Such conditions may arise, for example, in ultrasound imaging [8], atomic-resolution electron cryo-microscopy (cryo-EM) [14] and soft X-ray imaging [15].

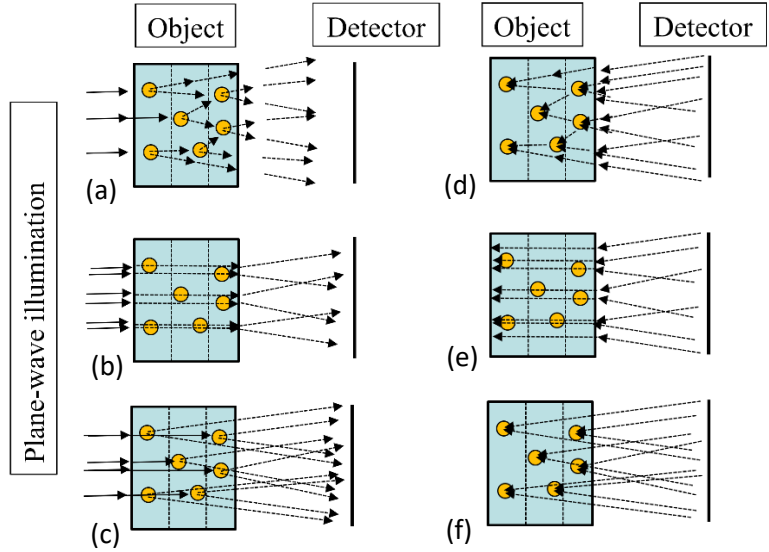

Fig. 1. Three variants of the forward imaging model and the corresponding variants of 3D object reconstruction (solution of the inverse imaging problem). (a) Full multislice model [11] of the in-object propagation: both the Fresnel diffraction and the (re)scattering are taken into account repeatedly in each transverse "slice"; multiple scattering is incorporated. (b) Projection approximation / straight ray model of the in-object propagation [5]: the Fresnel diffraction inside the object is not taken into account, (re)scattering is allowed to happen multiple times for each incident ray, but only along the straight line trajectories. (c) First Born / weak diffraction model of the in-object propagation [7]: the Fresnel diffraction inside the object is taken into account, but the scattering is allowed to happen no more than once for each incident ray; each (weak) scattering centre is considered illuminated by the unperturbed incident beam. All three models of in-object propagation are usually combined with the same (Fresnel) model of free-space propagation from the object to the detector. These models can be used for 3D reconstruction of an object from transmission images collected at multiple illumination directions. (d) Reconstruction using the full multislice model: 3D distribution of the refractive index inside the object is iteratively refined on the basis of forward multislice propagation through the reconstructed object, free-space propagation to the detector and comparison with the experimental images. (e) Conventional or phase-contrast CT. First, phase retrieval is applied to the experimental images and the free-space propagation from the object to the detector is numerically reversed, if necessary ("CTF correction"). Second, conventional CT is applied to reconstruct the 3D distribution of the refractive index inside the object. (f) Diffraction tomography: phase retrieval is applied to the experimental images and the free-space propagation from different planes inside the object to the detector is numerically reversed, resulting in the reconstruction of the 3D distribution of the refractive index inside the object.

Finally, the so-called phase-contrast CT (PCT) [16-20] occupies the ground approximately in between the conventional CT and the DT (Fig.1). In a typical PCT experiment, the phase shifts introduced by the imaged object into the transmitted beam are used explicitly to reconstruct the 3D spatial distribution of the refractive index inside the object. In this process, the Fresnel diffraction effects are taken into account, as in DT. However, unlike the DT, it is usually only the Fresnel diffraction in free space (outside of the object), which takes place during the propagation of the transmitted radiation from the object plane to the detector plane, that is taken into account in most PCT methods [21,22]. The in-object Fresnel diffraction and the corresponding Ewald sphere curvature are usually neglected in PCT. Accordingly, after numerical Fresnel back-propagation of the complex amplitude from the detector plane to the object plane at each illumination direction, the conventional CT model, based on the straight-line propagation of the probing radiation or matter waves inside the object, is used in PCT reconstructions. In contrast with the conventional CT, where only the 3D distribution of the imaginary (absorptive) part of the refractive index can be reconstructed, PCT can recover the 3D distribution of the real part of the complex refractive index, or, in some cases, the 3D distribution of both the real and the imaginary parts. Note that, for example, in hard X-ray imaging, the real decrement of the complex refractive index of soft biological tissues is about three orders of magnitude larger than the imaginary part, leading to higher PCT contrast in imaging of unstained samples compared to conventional CT [23]. For that reason, PCT has enjoyed widespread application in biomedical imaging in recent years [20,22].

Phase-contrast CT and DT imaging typically involve the so-called phase retrieval step, which is needed in order to recover the complex amplitude of the transmitted wave from its intensity registered by a position-sensitive detector in one or more planes orthogonal to the illumination direction [10,22]. Knowing the complex amplitude in a transverse plane allows one to numerically back-propagate the wave from the detector plane to the object plane in PCT, or to different planes inside the volume occupied by the imaged object in DT [24]. In electron microscopy, the phase retrieval step is often referred to as contrast transfer function (CTF) correction [14]. The CTF correction tends to be relatively straightforward in this context, because of the fact that atoms can be treated as pure phase objects when imaged with high-energy electrons, in the sense that scattering of an electron wave on an atom changes the phase of the wave, but not its intensity [11]. It is well known that phase retrieval in the case of pure phase objects is much simpler compared to the general case involving both phase shifts and absorption, and can be accomplished using the intensity data collected in a single image plane [22,25]. A useful generalization of the concept of pure phase objects is represented by the class of so-called "monomorphous" (also sometimes called "homogeneous") objects [22,26,27], for which the phase shift and the absorption are proportional to each other, with the coefficient of proportionality being the same at any position inside the object (see the next section for details). Similarly to the case of pure phase objects, when imaging monomorphous objects, the phase retrieval (CTF correction) can be accomplished using intensity data collected in a single transverse plane [25]. For general objects, in which the ratio of phase shifts (refraction) and absorption can be different in different parts of the object, phase retrieval generally requires intensity data to be collected in two or more different transverse planes or at two or more different radiation wavelengths [22,25]. While this can be accomplished in specially designed experiments, it typically necessitates higher radiation doses and higher requirements to the mechanical and optical stability compared to single-plane measurements. Therefore, in most practical applications, especially where minimization of the radiation dose is essential, as in biomedical imaging, or where sufficient stability of imaging conditions is difficult to achieve due to effects of radiation damage, as e.g. in cryo-EM, it is highly advantageous to use a 3D reconstruction method that requires just one transmission image per illumination direction. For this reason, the method considered in the present paper is focused on this type of imaging configuration.

## 2. Unified reconstruction method for conventional, phase-contrast and diffraction CT

Let a plane wave with complex amplitude $U_{in}(\mathbf{r}) = I_{in}^{1/2} \exp(i2\pi\lambda z)$ illuminate a thin object with transmission function $T(\mathbf{r}_\perp) = \exp[\psi_0(\mathbf{r}_\perp)]$, where $\psi_0(\mathbf{r}_\perp) = -a(\mathbf{r}_\perp) + i\varphi(\mathbf{r}_\perp)$. When the projection approximation [22] can be used to calculate the transmission function, one has $a(\mathbf{r}_\perp) = (2\pi/\lambda)\int \beta(\mathbf{r}_\perp, z)dz$ and $\varphi(\mathbf{r}_\perp) = (2\pi/\lambda)\int \delta(\mathbf{r}_\perp, z)dz$, where $n(\mathbf{r}) = 1 + \delta(\mathbf{r}) + i\beta(\mathbf{r})$ is the complex refractive index distribution inside the object (we omit the dependence of the refractive index on the wavelength for brevity). Note that $\beta(\mathbf{r})$ is always non-negative, while $\delta(\mathbf{r}) \leq 0$ in the case of hard X-ray imaging and $\delta(\mathbf{r}) = V(\mathbf{r})/(2E) \geq 0$ in TEM, where $V(\mathbf{r})$ is the electrostatic potential and $E$ is the accelerating voltage. The transmitted beam, $U(\mathbf{r}_\perp, z) = U_{in}(\mathbf{r})\exp[\psi(\mathbf{r}_\perp, z)]$, with $\psi(\mathbf{r}_\perp, 0) = \psi_0(\mathbf{r}_\perp)$, then propagates along the optic axis $z$ and a 2D image is collected in a plane $z = R$. This free-space propagation is described by the paraxial equation:

$$i(4\pi/\lambda)\partial_z U(\mathbf{r}_\perp, z) + \nabla^2_\perp U(\mathbf{r}_\perp, z) = 0. \quad (1)$$

Substituting $U(\mathbf{r}_\perp, z) = U_{in}(\mathbf{r})\exp[\psi(\mathbf{r}_\perp, z)]$ into eq.(1), we obtain the following equation for $\psi(\mathbf{r}_\perp, z)$:

$$i(4\pi/\lambda)\partial_z \psi(\mathbf{r}_\perp, z) + \nabla^2_\perp \psi(\mathbf{r}_\perp, z) + |\nabla_\perp \psi(\mathbf{r}_\perp, z)|^2 = 0. \quad (2)$$

It is now easy to see that, if $|\nabla_\perp \psi(\mathbf{r}_\perp, z)|^2 \ll |\nabla^2_\perp \psi(\mathbf{r}_\perp, z)|$ and so the term $|\nabla_\perp \psi(\mathbf{r}_\perp, z)|^2$ can be omitted from eq.(2) (which can happen, for example, when $|\psi(\mathbf{r}_\perp, z)| \ll 1$), the function $\psi(\mathbf{r}_\perp, z)$ itself satisfies the paraxial eq.(1). This represents a form of Rytov approximation in free space [22,28]. A solution to eq.(1) is given by the usual Fresnel diffraction integral [11], which can be expressed in the reciprocal space as a product of the Fresnel propagator with the Fourier transform, $(\mathbf{F}_2\psi_0)(\mathbf{q}_\perp) = \int \exp(-i2\pi\mathbf{r}_\perp \cdot \mathbf{q}_\perp)\psi_0(\mathbf{r}_\perp)d\mathbf{r}_\perp$, of the initial function in the object plane:

$$(\mathbf{F}_2\psi)(\mathbf{q}_\perp, R) = \exp(-i\pi\lambda R q^2_\perp)(\mathbf{F}_2\psi_0)(\mathbf{q}_\perp). \quad (3)$$

Adding together eq.(3) and a similar equation for the complex conjugate of $\psi$, $(\mathbf{F}_2\psi^*)(\mathbf{q}_\perp, R) = \exp(i\pi\lambda R q^2_\perp)(\mathbf{F}_2\psi_0^*)(\mathbf{q}_\perp)$, we find that the function $2(\mathbf{F}_2 \mathrm{Re}\psi)(\mathbf{q}_\perp, R) = \mathbf{F}_2[\ln(I/I_{in})](\mathbf{q}_\perp, R)$ satisfies the so-called CTF equation:

$$\mathbf{F}_2[\ln(I/I_{in})](\mathbf{q}_\perp, R) = 2(\mathbf{F}_2\varphi)(\mathbf{q}_\perp)\sin(\pi\lambda R q^2_\perp)] - 2(\mathbf{F}_2 a)(\mathbf{q}_\perp)\cos(\pi\lambda R q^2_\perp). \quad (4)$$

Note that replacing $\ln(I/I_{in}) \cong I/I_{in} - 1$ in the left-hand side of eq.(4) leads to the well-known first Born (weak object) approximation for the propagated intensity [11,29].

Now consider a model in which the imaged object is "split" into $M$ thin transverse slices (slabs) located one next to another at $z$-positions $z_m$, $m = 1,...,M$, as in the multi-slice approximation [11]. The intensity produced by each slice in isolation satisfies eq.(4) with $\varphi(\mathbf{r}_\perp, z_m) = (2\pi/\lambda)\int_{z_{m-1}}^{z_m} \delta(\mathbf{r}_\perp, z)dz \cong (2\pi/\lambda)\delta(\mathbf{r}_\perp, z_m)\Delta z$ and $a(\mathbf{r}_\perp, z_m) = (2\pi/\lambda)\int_{z_{m-1}}^{z_m} \beta(\mathbf{r}_\perp, z)dz \cong (2\pi/\lambda)\beta(\mathbf{r}_\perp, z_m)\Delta z$, where $\Delta z$ is the slice thickness. Following the approach used in [30], we approximate the total image intensity distribution produced by the whole object by the incoherent sum of contributions from the individual thin slices:

$$\mathbf{F}_2[\ln(I/I_{in})](\mathbf{q}_\perp, R) = (4\pi/\lambda) \\ \times \sum_{m=1}^{M}\{(\mathbf{F}_2\delta)(\mathbf{q}_\perp, z_m)\sin[\pi\lambda(R-z_m)q^2_\perp] - (\mathbf{F}_2\beta)(\mathbf{q}_\perp, z_m)\cos[\pi\lambda(R-z_m)q^2_\perp]\}\Delta z. \quad (5)$$

Increasing the number of slices, $M \to \infty$, and simultaneously making them thinner, $\Delta z \to 0$, so that the total object thickness $M\Delta z$ remains the same, we can derive an integral form of eq.(5):

$$\mathbf{F}_2[\ln(I/I_{in})](\mathbf{q}_\perp, R) = (4\pi/\lambda) \\ \times \int\{(\mathbf{F}_2\delta)(\mathbf{q}_\perp, z)\sin[\pi\lambda(R-z)q^2_\perp] - (\mathbf{F}_2\beta)(\mathbf{q}_\perp, z)\cos[\pi\lambda(R-z)q^2_\perp]\}dz. \quad (6)$$

Equation (6) can serve as a basis for a Diffraction Tomography (DT) type object reconstruction formalism [7-10, 30], which is further developed below.

As we are ultimately interested in a reconstruction method that can use a single transmission image at each illumination direction, we now introduce a class of "monomorphous" (alternatively called "homogeneous") objects which satisfy the condition $\beta(\mathbf{r}) = \sigma\delta(\mathbf{r})$, with a constant $\sigma$ [25,26,31] ($\sigma \geq 0$ in TEM and $\sigma \leq 0$ in hard X-ray imaging). Note that the sub-classes of "pure-phase" and "pure-absorption" objects formally correspond here to the cases $\sigma = 0$ and $|\sigma| = \infty$, respectively. For monomorphous objects, eq.(6) becomes

$$\mathbf{F}_2[\ln(I/I_{in})](\mathbf{q}_\perp, R) = (4\pi\sqrt{1+\sigma^2}/\lambda) \int \sin[\pi\lambda(R-z)q_\perp^2 - \omega](\mathbf{F}_2\delta)(\mathbf{q}_\perp, z)dz, \tag{7}$$

where $\omega = \tan^{-1}(\sigma)$, and hence $\sin\alpha - \sigma\cos\alpha = \sqrt{1+\sigma^2}\sin(\alpha - \omega)$ for any $\alpha$. Using the identity

$$\sin[\pi\lambda(R-z)q_\perp^2 - \omega] = \{\exp[i\pi\lambda(R-z)q_\perp^2 - i\omega] - \exp[-i\pi\lambda(R-z)q_\perp^2 + i\omega]\}/(2i)$$
$$= \{\exp(-i2\pi z\lambda q_\perp^2/2)\exp[i(\pi\lambda Rq_\perp^2 - \omega)] - \exp(i2\pi z\lambda q_\perp^2/2)\exp[i(-\pi\lambda Rq_\perp^2 + \omega)]\}/(2i),$$

eq.(7) can be re-written as

$$\mathbf{F}_2[\ln(I/I_{in})](\mathbf{q}_\perp, R) = (i2\pi\sqrt{1+\sigma^2}/\lambda)$$
$$\times \{\exp[-i(\pi\lambda Rq_\perp^2 - \omega)](\mathbf{F}_3\delta)(\mathbf{q}_\perp, -\lambda q_\perp^2/2) - \exp[i(\pi\lambda Rq_\perp^2 - \omega)](\mathbf{F}_3\delta)(\mathbf{q}_\perp, \lambda q_\perp^2/2)\}. \tag{8}$$

At the opposite illumination direction, i.e. for the image, $I_\pi(\mathbf{r}_\perp, R)$, of the object rotated by 180 degrees around the $y$ axis, we have [30]:

$$\mathbf{F}_2[\ln(I_\pi/I_{in})](\mathbf{q}_\perp^-, R) = (4\pi/\lambda)\int\{\sin[\pi\lambda(R-z)q_\perp^2] - \sigma\cos[\pi\lambda(R-z)q_\perp^2]\}(\mathbf{F}_2\delta)(\mathbf{q}_\perp, -z)dz$$
$$= (4\pi/\lambda)\int\{\sin[\pi\lambda(R+z)q_\perp^2] - \sigma\cos[\pi\lambda(R+z)q_\perp^2]\}(\mathbf{F}_2\delta)(\mathbf{q}_\perp, z)dz$$
$$= (4\pi\sqrt{1+\sigma^2}/\lambda)\int \sin[\pi\lambda(R+z)q_\perp^2 - \omega](\mathbf{F}_2\delta)(\mathbf{q}_\perp, z)dz,$$

where $\mathbf{q}_\perp^- \equiv (-q_x, q_y)$ is the mirror-reflection of the vector $\mathbf{q}_\perp = (q_x, q_y)$ with respect to the axis of rotation. Using another identity,

$$\sin[\pi\lambda(R+z)q_\perp^2 - \omega] = \{\exp[i\pi\lambda(R+z)q_\perp^2 - i\omega] - \exp[-i\pi\lambda(R+z)q_\perp^2 + i\omega]\}/(2i)$$
$$= \{\exp(i2\pi z\lambda q_\perp^2/2)\exp[i(\pi\lambda Rq_\perp^2 - \omega)] - \exp(-i2\pi z\lambda q_\perp^2/2)\exp[-i(\pi\lambda Rq_\perp^2 - \omega)]\}/(2i),$$

we can now obtain:

$$\mathbf{F}_2[\ln(I_\pi/I_{in})](\mathbf{q}_\perp^-, R) = (i2\pi\sqrt{1+\sigma^2}/\lambda)$$
$$\times \{\exp[-i(\pi\lambda Rq_\perp^2 - \omega)](\mathbf{F}_3\delta)(\mathbf{q}_\perp, \lambda q_\perp^2/2) - \exp[i(\pi\lambda Rq_\perp^2 - \omega)](\mathbf{F}_3\delta)(\mathbf{q}_\perp, -\lambda q_\perp^2/2)]. \tag{9}$$

Solving the system of linear eqs. (8) and (9) for the refractive index, we arrive at the following explicit expression for the 3D distribution of $\delta$ as a function of 2D images:

$$(\mathbf{F}_3\delta)(\mathbf{q}_\perp, -\lambda q_\perp^2/2) = \frac{\lambda}{4\pi\sqrt{1+\sigma^2}\sin[2(\pi\lambda Rq_\perp^2 - \omega)]}\Big[\exp[-i(\pi\lambda Rq_\perp^2 - \omega)]$$
$$\times \mathbf{F}_2[\ln(I/I_{in})](\mathbf{q}_\perp, R) + \exp[i(\pi\lambda Rq_\perp^2 - \omega)]\mathbf{F}_2[\ln(I_\pi/I_{in})](\mathbf{q}_\perp^-, R)\Big]. \tag{10}$$

Equation (10) is our key result; it represents a form of the "Fourier diffraction projection theorem" [32] and provides a DT solution for reconstruction of 3D distributions of the refractive index in pure-phase, pure-absorption and monomorphous objects from 2D transmission images collected at different rotational positions of the object. This equation generalises our previous result, expressed by eq.(6) in [30]. Related forms of "intensity-only DT" have been previously investigated and implemented [10,21] . Note that eq.(10) takes into account the curvature of the Ewald sphere or, equivalently, takes into account the Fresnel diffraction inside the object. In order to be able to reconstruct the 3D distribution of $\delta$ from a set of 2D transmission images using eq.(10), the set of corresponding illumination directions should be sufficiently "rich" to allow for a proper sampling of all Fourier coefficients $(\mathbf{F}_3\delta)(\mathbf{q})$ within the chosen reconstruction volume at the chosen spatial resolution via the rotation of the paraboloid $(\mathbf{q}_\perp, -\lambda q_\perp^2 / 2)$ over the available illumination directions. We will show below that eq.(10) includes phase-contrast CT [22] and conventional CT [5] as special cases.

For a practical tomographic reconstruction from multiple images collected at different illumination directions, eq.(10) can be re-written in the following way:

$$(\mathbf{F}_3\delta)\left(\mathbf{q}_\perp, -\lambda q_\perp^2 / 2\right) \cong \frac{\lambda}{4\pi n_{\mathbf{q}_\perp} \sqrt{1+\sigma^2}} \sum_{n=1}^{n_{\mathbf{q}_\perp}} \mathbf{F}_2[\ln(I_n / I_{in})](\mathbf{q}_\perp, R_n)$$
$$\times \left[ \frac{\sin(\pi\lambda R_n q_\perp^2 - \omega)}{\sin^2(\pi\lambda R_n q_\perp^2 - \omega) + \varepsilon} - i \frac{\cos(\pi\lambda R_n q_\perp^2 - \omega)}{\cos^2(\pi\lambda R_n q_\perp^2 - \omega) + \varepsilon} \right],$$
(11)

where the index $n$ spans the subset of available illumination directions that are orthogonal to $\mathbf{q}_\perp$, $R_n$ are the corresponding positions of the image planes and $\varepsilon$, $0 < \varepsilon \ll 1$, is the usual Tikhonov regularization parameter preventing a division by zero. Note that eq.(11), in principle, still relies on the availability of projections $I_n$ at (at least, approximately) opposite directions in the input dataset, so that the sum in eq.(11) contains contributions from such pairs of projections, as required by eq.(10). In practice, a 3D gridding can be applied to calculate the contribution from each image contrast function $\ln(I_n / I_{in})$ to the reconstructed distribution of $\delta$ [33,34]. In that case, the factors $n_{\mathbf{q}_\perp}^{-1}$ in eq.(11) are replaced by the (regularized) inverse values, $S_\varepsilon^{-1}(\mathbf{q})$, of the "sampling matrix", whose elements $S(\mathbf{q})$ are the sums of all fractional (gridding) weights corresponding to contributions of different contrast functions, $\mathbf{F}_2[\ln(I_n / I_{in})](\mathbf{q}_\perp, R_n)$, to a given Fourier coefficient $(\mathbf{F}_3\delta)(\mathbf{q})$. Conceptually, the coefficients $S(\mathbf{q})$ are similar to the weights in a 3D linear interpolation. Each "reciprocal pixel" value $\mathbf{F}_2[\ln(I_n / I_{in})](\mathbf{q}_\perp, R_n)$ contributes to exactly eight adjacent values $(\mathbf{F}_3\delta)(\mathbf{q})$ at the location of eight vertices of the cube with a fixed side length (equal to the spatial resolution of the reconstruction) containing the point $(\mathbf{q}_\perp, R_n)$. The respective weights of the eight contributions are inversely proportional to the distance between the point $(\mathbf{q}_\perp, R_n)$ and the corresponding vertices, with the sum of all eight weights equal to one. This type of direct Fourier space gridding-based reconstruction can accommodate, in principle, any tomographic scanning geometry (regular or random) with plane incident waves. In order to obtain a good-quality reconstruction, the usual CT sampling conditions must be satisfied [5,35]. As an example, in the case of conventional planar CT scanning with uniform rotational steps over 180 degrees around a fixed Y axis, the sampling coefficients naturally approximate the well-known ramp filter, $S(\mathbf{q}) = |q_x|$, in accordance with the equidistant sampling density in the cylindrical coordinates in 3D space [5]. This simple 3D gridding implementation of eq.(11), which we call the Unified Tomographic Reconstruction (UTR) algorithm, has been realised in our publicly-released software [36]. Examples of 3D reconstruction from electron and X-ray CT data using the UTR algorithm are given in the next section.

Multiple scattering that may take place during the propagation of radiation or matter waves through an imaged object has not been explicitly taken into account in eq.(11). However, in the presence of significant multiple scattering, an approach to 3D reconstruction from transmission images based on eq.(11) may still work well due to "averaging out" of multiple scattering effects in the process of summation of contributions from images collected at different incident directions. In the majority of real situations, multiple scattering tends to be strongly anisotropic and its contributions tend to blur into a quasi-uniform background when averaged over many different orientations [30]. The background can then be thresholded or low-pass filtered out of the reconstruction. The effectiveness of such an approach can be observed, in particular, in a simulated example of the application of eq.(11) to 3D reconstruction of a Fe-Pt nanoparticle in the next section of this paper, which involves a significant amount of multiple scattering. In the cases where multiple scattering effects on the 3D reconstruction cannot be satisfactory handled by simple angular averaging and thresholding, it may still be necessary to use more sophisticated methods and incorporate multiple scattering explicitly into the 3D reconstruction framework [37,38].

Similarly to the results demonstrated recently in [39,40], when the defocus distances at different orientations are the same, $R_n = R$, the phase retrieval and the CT reconstruction steps in eq.(11) can be swapped, so that the CT reconstruction is performed first and the phase retrieval is applied subsequently in 3D:

$$(\mathbf{F}_3 \delta)\left(\mathbf{q}_\perp, -\lambda q_\perp^2 / 2\right) \cong \left[ \frac{\sin(\pi \lambda R q^2 - \omega)}{\sin^2(\pi \lambda R q^2 - \omega) + \varepsilon} - i \frac{\cos(\pi \lambda R q^2 - \omega)}{\cos^2(\pi \lambda R q^2 - \omega) + \varepsilon} \right]$$

$$\times \frac{\lambda}{4\pi n_{\mathbf{q}_\perp} \sqrt{1+\sigma^2}} \sum_{n=1}^{n_{\mathbf{q}_\perp}} \mathbf{F}_2[\ln(I_n / I_{in})](\mathbf{q}_\perp, R). \tag{12}$$

The latter form of the UTR algorithm can be beneficial in certain types of applications [39], where repeated phase retrieval with different values of parameter $\omega$ may be required. In such situations, eq.(12) allows one to avoid re-running the computationally-expensive 3D CT reconstruction step. However, the variant expressed by eq.(11) is generally more flexible with respect to the experimental conditions than eq.(12). In particular, it is straightforward to introduce optical aberrations and other imaging conditions in eq.(11) that can vary between different views, simply by modifying the arguments of the sine and cosine terms of the CTF [11]. Such additional optional modifications may be relevant, for example, in single-particle analysis in cryo-EM [14], they have been implemented in our software [36], which includes fast implementations of both eqs.(11) and (12).

The Fourier slice theorem of phase-contrast CT of monomorphous objects (including pure phase-contrast CT) is directly obtained from eq.(10) when the Ewald sphere curvature is so small that the term $(\lambda/2)q_\perp^2$ in the left-hand side of eq.(10) can be replaced by zero. It is easy to see from eq.(7) that a sufficient condition for this is $\lambda z q_\perp^2 \ll 1$ at all $z$ inside the object, or $\Delta^2 / (T\lambda) \gg 1$, where $\Delta = 1/|q_{max}|$ is the transverse spatial resolution and $T$ is the thickness of the imaged object. The latter condition indicates that the DOF is much larger than the object thickness. It can be seen from eqs.(8) and (9) that, when $(\lambda/2)q_\perp^2$ is replaced by zero, the images collected at the opposite illumination directions are identical, that is $I_\pi(\mathbf{r}_\perp^-, R) = I(\mathbf{r}_\perp, R)$. Under such conditions, eq.(10) becomes:

$$(\mathbf{F}_3 \delta)(\mathbf{q}_\perp, 0) = \frac{\lambda}{4\pi\sqrt{1+\sigma^2}} \mathbf{F}_2[\ln(I / I_{in})](\mathbf{q}_\perp, R)$$

$$\times \left[ \frac{\exp[-i(\pi\lambda R q_\perp^2 - \omega)] + \exp[i(\pi\lambda R q_\perp^2 - \omega)]}{\sin[2(\pi\lambda R q_\perp^2 - \omega)]} \right] = \frac{\lambda \mathbf{F}_2[\ln(I/I_{in})](\mathbf{q}_\perp, R)}{4\pi\sqrt{1+\sigma^2} \sin(\pi\lambda R q_\perp^2 - \omega)}. \tag{13}$$

Note that, because the images at the opposite orientations are identical here, eq.(13) no longer requires pairs of images collected at opposite orientations to be available in the CT input dataset. Let us verify that the last equation coincides with the Fourier slice theorem for CT of monomorphous objects (including pure-phase objects). The equation for the image intensity in the case of a CT based on the projection approximation of weak monomorphous objects is $\mathbf{F}_2[\ln(I/I_{in})](\mathbf{q}_\perp, R) = 2\sqrt{1+\sigma^2}\sin(\pi\lambda R q_\perp^2 - \omega)(\mathbf{F}_2\varphi_0)(\mathbf{q}_\perp)$ [25], where $\varphi_0(\mathbf{r}_\perp) = (2\pi/\lambda)\int \delta(\mathbf{r}_\perp, z)dz$ is the projected phase (see also eq.(4) above in the monomorphous case). In particular, for pure phase objects one has $\sigma = \omega = 0$ and hence $\mathbf{F}_2[\ln(I/I_{in})](\mathbf{q}_\perp, R) = 2\sin(\pi\lambda R q_\perp^2)(\mathbf{F}_2\varphi_0)(\mathbf{q}_\perp)$ [11]. Substituting this into the right-hand side in the second line of eq.(13), we obtain that $(\mathbf{F}_3\delta)(\mathbf{q}_\perp, 0) = [\lambda/(2\pi)](\mathbf{F}_2\varphi_0)(\mathbf{q}_\perp)$. The last equation is indeed the Fourier slice theorem of phase CT [5,22]. It can be obtained simply by 2D Fourier transforming the equation for the phase function, $\varphi_0(\mathbf{r}_\perp) = (2\pi/\lambda)\int \delta(\mathbf{r}_\perp, z)dz$. Note that this Fourier slice theorem holds for strong pure phase and monomorphous objects, provided that the projection approximation holds for the transmission of the incident wave through the object [25].

The same result (eq.(13)) can be directly obtained from eq.(11) with $n_{\mathbf{q}_\perp} = 1$, if we discard the second additive term (i.e. the imaginary part) in the square brackets on the right-hand side of eq.(11). The disappearance of the imaginary part here corresponds to the fact that, when the Ewald sphere curvature is negligible and $I_\pi(\mathbf{r}_\perp^-, R) = I(\mathbf{r}_\perp, R)$, the imaginary parts of the square brackets in eq.(11) at any two opposite orientations will have the same absolute value and opposite signs and hence will cancel each other. Therefore, when the Ewald sphere curvature is negligible, eq.(11) naturally transforms into the conventional CT reconstruction of monomorphous objects based on the rotation of the central slice in the Fourier space [5].

If the propagation distance $R$ is small, corresponding to imaging in the near-Fresnel region, where $\lambda R q_\perp^2 \ll 1$ or $\lambda R/\Delta^2 \ll 1$, the sine term in the denominator of eq.(13) can be approximated as $\sin(\pi\lambda R q_\perp^2 - \omega) = \sin(\pi\lambda R q_\perp^2)\cos\omega - \cos(\pi\lambda R q_\perp^2)\sin\omega \cong (\pi\lambda R q_\perp^2 - \sigma)/\sqrt{1+\sigma^2}$, where we have used the identities $\cos\omega = 1/\sqrt{1+\sigma^2}$ and $\sin\omega = \sigma/\sqrt{1+\sigma^2}$. This converts eq.(13) into the so-called homogeneous Transport of Intensity equation [26] in the Fourier space representation:

$$(\mathbf{F}_3\mu)(\mathbf{q}_\perp, 0) = -\frac{\mathbf{F}_2[\ln(I/I_{in})](\mathbf{q}_\perp, R)}{1 - (\pi\lambda R/\sigma)q_\perp^2}. \qquad (14)$$

where $\mu = (4\pi/\lambda)\beta = (4\pi/\lambda)\sigma\delta$ is the linear attenuation coefficient. As we mentioned previously, $\sigma \leq 0$ in hard X-ray imaging. Therefore the denominator of eq.(14) is always separated from zero in the latter context, making eq.(14) highly stable with respect to noise in the input images [22].

When the propagation distance $R$ becomes negligibly small and can be replaced by zero in eq.(14), the latter equation morphs into the Fourier slice theorem of conventional (absorption-based) CT [5]:

$$(\mathbf{F}_3\mu)(\mathbf{q}_\perp, 0) = -\mathbf{F}_2[\ln(I/I_{in})](\mathbf{q}_\perp, 0). \qquad (15)$$

Note again that this Fourier slice theorem holds for strongly absorbing objects, regardless of the phase distribution, as long as the projection approximation holds for the transmission of the incident wave through the object [5].

## 3. Reconstruction examples with simulated and experimental images

The proposed UTR algorithm for 3D CT reconstruction is based on eqs.(11) and (12). As mentioned above, it uses three-dimensional gridding [34] for calculating the contribution of all the projections available at different orientations to the reconstructed 3D distribution of the refractive index. Because of the universality of this Fourier space gridding approach, the UTR method can be used with image data obtained in any scanning geometry involving plane-wave illumination. The spatial filter function which is typically required in a CT reconstruction is generated here automatically inside the program on the basis of the actual sampling of the input data. The reconstruction examples presented below have been obtained using a C++ implementation of the UTR algorithm [36]. The input data used in these examples can be obtained upon a request sent by email to the first author of this paper.

In the first example, we reconstruct a nanoparticle containing 5,107 Pt atoms and 5,356 Fe atoms [41], placed on a 100 Å thick amorphous carbon substrate containing 90,253 C atoms. We used the same test object in our recent publication [24], where it was reconstructed using different methods. Working with the same test object here allows us to quantitatively compare the performance of the UTR algorithm with the best results obtained with previously published methods, including the vCTF method [24], DHT and CHR methods [42] and the conventional CTF-corrected CT. The numerical simulations reported here have been performed using open-source software [36,43]. For the simulations, the test object consisting of the nanoparticle and the substrate was placed into a virtual cubic volume with 200 Å sides located in the positive octant of the Cartesian coordinates in 3D, with the sides parallel to the coordinate axes and one corner located at the origin of coordinates. A 3D rendering of the nanoparticle and the substrate in one of the used orientations is shown in Fig.2(a). The 3D image was obtained using the Vesta software [44] with the input file containing the 3D coordinates of all atoms. The structure was illuminated by a plane monochromatic electron beam with E = 200 keV ( $\lambda \cong 0.025\,\text{Å}$ ) propagating along the $z$ axis. The simulated scan consisted of images collected at 360 different pseudo-random orientations of the particle with the substrate, with the directional vectors of the orientations uniformly distributed on the unit sphere in 3D. The effective centre of rotation was at the centre of the virtual cube containing the object, i.e. at the point $(x, y, z) = (100\,\text{Å}, 100\,\text{Å}, 100\,\text{Å})$. The defocused images with $1{,}024 \times 1{,}024$ pixels each were obtained by first calculating the transmission of the electron wave through the object using a multislice-based algorithm [45-47] and then calculating the free-space propagation from the exit plane of the object to the detector by computing the appropriate Fresnel diffraction integrals. In these calculations, an effective objective aperture of 40 mrad was assumed, achievable in aberration-corrected TEM. The effect of thermal motion of atoms was included in the simulations via a Debye-Waller factor with the root-mean square displacement of 0.085 Å at 300 K. For different orientations of the object, the image planes were located at different positions which were uniformly randomly distributed between $z = 300$ Å and $z = 350$ Å. Subsequently, we "added" pseudo-random Poisson shot noise with a mean corresponding to 59 electrons per Å$^2$ (approximately 2.25 electrons per pixel) to each image. A typical simulated noisy 2D defocused image at one of the 3D orientations is shown in Fig.2(b).

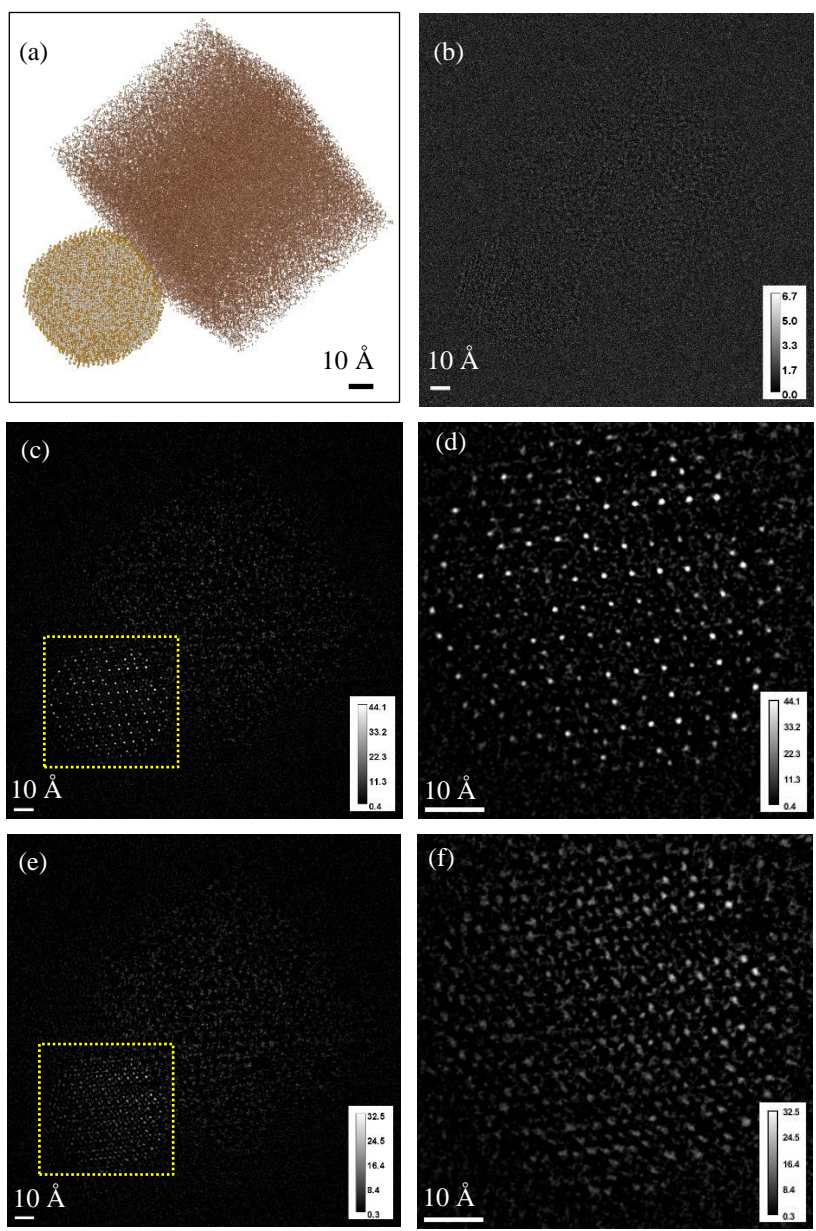

Fig. 2. (a) 3D rendering of the Fe-Pt nanoparticle on a C substrate used in the simulations included in section 3. (b) A typical simulated defocused image with Poisson noise. (c) 2D cross-section at the level $z = 100$ Å through the 3D distribution of the electrostatic potential $V(\mathbf{r}) = 2E\delta(\mathbf{r})$ reconstructed using the UTR method. (d) Zoomed (×3) version of the area shown by the yellow rectangle in (c). (e) 2D cross-section at $z = 100$ Å through the potential reconstructed from the same input images without the Ewald sphere curvature correction. (f) Zoomed (×3) version of the area shown by yellow rectangle in (e). The grey-scale levels in (c)-(f) are in volts.

We then applied the UTR algorithm, as implemented in [36], to the input set of 360 simulated noisy defocused images. The only two tuneable technical input parameters in this reconstruction were the mean noise-to-signal (NSR) ratio of 0.66667 (corresponding to the Poisson noise with the standard deviation of $\sqrt{2.25} = 1.5$, giving NSR = 1 / 1.5 $\cong$ 0.66667) and the Tikhonov regularization parameter $\varepsilon = 0.1$ (see the denominator of the terms in the square brackets on the right-hand side of eq.(11)). The value of the last parameter was the result of an ad-hoc "optimization" on the basis of previous experience and a few trials. As a general rule, this parameter should be small when the noise level in the input data is low, or it can be relatively large (as in the current example), when the input images are very noisy. The precise value of this parameter has a relatively weak effect on the result, with the reconstruction generally becoming smoother and blurrier as $\varepsilon$ increases, or sharper and noisier when $\varepsilon$ decreases.

Figure 2(c) shows one 2D slice through the UTR reconstruction of the 3D distribution of the electrostatic potential obtained from the set of 360 noisy defocused images as described above. Figure 2(d) contains a zoomed ($\times 3$) image of the region in Fig.2(c) containing the nanoparticle. Figure 2(e) and 2(f) show the corresponding images obtained from the same input images using the same reconstruction method, but with the Ewald sphere curvature correction "switched off". The latter corresponded to discarding the imaginary part in the square brackets in the right-hand side of eq.(11) and replacing $\lambda q_\perp^2 / 2$ with zero in the right-hand side. It can be seen from these images that the incorporation of the Ewald sphere curvature correction in this example resulted in a much better quality reconstruction. Note that the Ewald sphere curvature was significant in this case, since the DOF was approximately 8 Å, i.e. much smaller than the linear size of approximately 70 Å of the nanoparticle.

We also carried out a quantitative analysis of the reconstruction by localizing the individual atoms in the reconstructed 3D image of the nano-particle. This was done by dividing the reconstruction volume into non-overlapping cubes with the side length of 1.7 Å and finding the peaks (highest pixel values) inside each such box. In the case of the UTR result (with the Ewald sphere curvature correction), this allowed us to correctly find 10,394 out of 10,463 actual atomic positions in the Fe-Pt nanoparticle, with the average error of 0.13 Å and the maximum distance of 0.63 Å between the original and the reconstructed atomic locations. The result also contained 69 "false positive" findings, i.e. the peaks in the reconstructed potential that were not associated with any actual atoms in the particle. All Pt atoms were correctly located, while 69 Fe atoms were missed in this reconstruction ("false negatives"). Of the first 5,107 highest identified peaks in the reconstructed electrostatic potential, 4,861 corresponded to locations of Pt atoms in the original nanoparticle and only 246 of such peaks erroneously corresponded to the original locations of Fe atoms.

In the case of the reconstruction without the Ewald sphere curvature correction, only 7,452 atoms were correctly located using the same procedure as above, with 3,011 false positive localizations. The average distance between the identified peaks of the electrostatic potential and the actual atomic locations was 0.47 Å, and the maximum distance was 1.0 Å, the latter corresponded to the set maximum distance in the location matching program. The inferior quality of this reconstruction, compared to the previous one, can be explained by the effect of the Ewald sphere curvature on the reconstruction of the electrostatic potential around the atoms located more than one DOF away from the effective centre of rotation in the scan. It was demonstrated in [25,43] that the reconstructed atomic potential around such peripheral atoms gets blurred and suppressed due to the error in the effective back-propagation distance from the image plane to the scatterer (atom), the error occurring when the Ewald sphere curvature is not taken into account. Indeed, it is possible to see in Fig.2(f) that the atomic potential was particularly poorly reconstructed in the vicinity of atoms located at the largest distances from the centre of rotation.

We have also compared the reconstruction of the Fe-Pt nanoparticle produced by the UTR method with best result obtained using the previously published vCTF method [24]. It was found that the UTR reconstruction had a slightly better overall visual quality and higher spatial resolution compared to the vCTF reconstruction from the same input images. On the quantitative level, vCTF reconstruction allowed for the correct localization of 10,373 out of 10,463 actual atomic positions in the Fe-Pt nanoparticle, with the average error of 0.12 Å and the maximum distance of 0.88 Å between the original and the reconstructed atomic locations. These results were slightly inferior to the UTR results presented above. However, the "specificity" of the vCTF method was slightly better than that of the UTR method, i.e. the UTR reconstruction allowed for a better separation of the Pt atoms from the Fe atoms in the nanoparticle on the basis of the heights of the corresponding peaks of the reconstructed electrostatic potential. This issue may be worth investigating further in a future study. The UTR algorithm was also more than 10 times faster in terms of computing time. The improvement of the computational speed compared to the vCTF method was primarily due to the fact that eq.(11) effectively back-propagates each input image onto the corresponding patch of the 2D Ewald sphere, while vCTF and related methods, such as CHR and DHT [42], back-propagate each input image onto all 2D planes orthogonal to the illumination direction inside the reconstruction volume containing the object, with the longitudinal distance between these planes equal to the spatial resolution of the reconstruction. Note also that the vCTF result described above was also better than the best reconstructions that we were able to obtain previously with the DHT and CHR methods [42]. Therefore, at least in the case of this example, the UTR has outperformed all of these methods.

In the second example, we used an experimental phase-contrast CT dataset that was collected at the Imaging and Medical Beamline (IMBL) of the Australian Synchrotron. This dataset was obtained with a breast mastectomy sample in the course of researching the options for a future synchrotron-based breast cancer facility. Additional details about the experimental setup and the project in general can be found, for example, in [48]. The imaging experiment was conducted under a Human Ethics Certificate of Approval and with written consent from the patient to image the clinical specimen. The mastectomy sample was imaged in a complete, intact, unfixed state at IMBL using the propagation-based phase-contrast CT technique [48]. The professional pathology analysis of this mastectomy sample later reported invasive carcinoma of no special type with high-grade ductal carcinoma in situ and extensive microcalcifications. Some of the cancer lesions and microcalcifications can be seen in our PCT reconstructions. For the scan, the sample was placed in a thin-walled plastic cylinder with a diameter of 11 cm. The scan was collected at a clinically-relevant 4 mGy mean glandular dose, which was distributed evenly between 4,800 projections with a uniform angular step of 0.0375 degrees over 180 degrees. The scan was carried out using quasi-plane monochromatic X-rays with energy E = 32 keV and a free-space propagation distance of 6 m between the sample and the detector. The X-ray detector had a pixel size of $100 \times 100$ μm$^2$ and field of view 12.16 cm (horizontal) $\times$ 12.32 cm (vertical). A spatial resolution of approximately 170 μm (horizontal) $\times$ 150 μm (vertical) was measured in this imaging setup. This resolution was determined predominantly by the detector's point-spread function, while the slight asymmetry of the resolution was due to the anisotropic X-ray source at IMBL.

An example of a typical experimental PCT projection of the mastectomy sample is shown in Fig.3(a). Figure 3(b) contains a 2D slice through the central plane of the reconstructed distribution of the scaled imaginary part, $\beta(\mathbf{r}) \times 10^{11}$, of the complex refractive index inside the mastectomy sample obtained from the 4,800 experimental PBI projections using the UTR method. Light-grey areas in this image correspond to denser glandular tissue and tumour regions, while darker grey areas correspond to the less dense adipose tissue. The thin vertical lines in the middle of the image are artefacts due to non-linear response of a few "bad" pixels in the detector, which could not be completely corrected during the standard image pre-processing. Note that the pre-processing included, in particular, the division of the "raw" projections by the "flat field" images collected under the same illumination conditions, but without the sample in place. Figure 3(c) contains a 2D slice at the same position as in Fig.3(b), but taken through the reconstructed distribution of $\beta(\mathbf{r}) \times 10^{11}$ obtained using the UTR method from the subset of only 480 PBI projections with a uniform angular step of 0.375 degrees over 180 degrees. The effective radiation dose corresponding to this sub-scan was 1/10th of the original one, i.e. only 0.4 mGy. Note that it is highly desirable to reduce the radiation dose delivered to the patient during an X-ray CT scan, since the breast is the most radio-sensitive part of the human body [49]. One can see that, because of the reduced mean photon fluence, the image in Fig.3(c) is noisier than the one in Fig.3(b). However, most of the important features of the sample are still clearly discernable in the low-dose image. We plan to perform a systematic radiological evaluation of similar low-dose breast PCT images in the near future in order to find out if such images could be acceptable for reliable cancer diagnosis. In this context, it is important to compare Fig.3(c) with Fig.3(d), the latter one representing the best result that we were able to obtain from the same 480 projections using the standard Filtered Back-Projection (FBP) method implemented in our well-tested and optimized PCT reconstruction software, X-TRACT [50]. The FBP-reconstructed image in Fig.3(d) is considerably noisier compared to the UTR image in Fig.3(c), which implies a potential for reduction of the radiation dose using the UTR method for the PCT reconstructions.

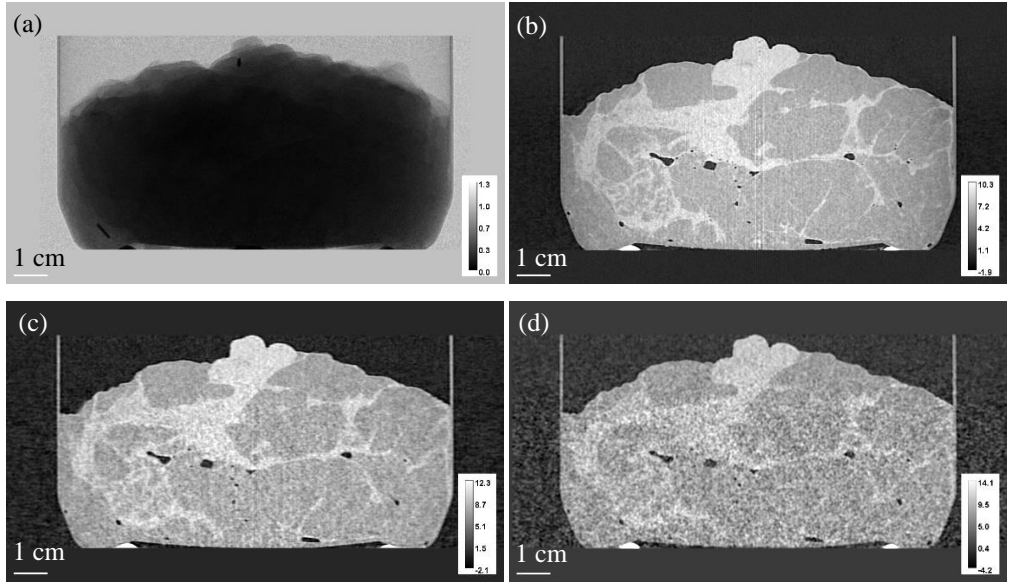

Fig. 3 (a) A typical experimental X-ray projection of the mastectomy sample collected at E = 32 keV, 0.84 μGy dose and R = 6 m. (b) 2D cross-section through the middle of the reconstructed 3D distribution of $\beta \times 10^{11}$ obtained from 4,800 projections using the UTR method. (c) Same as (b), but obtained from only 480 projections. (d) Same as (c), but obtained with the conventional FBP method.

Note that in the case of the second example above, the Ewald sphere curvature was clearly irrelevant, since here we had DOF ≅ 29 m, which was much larger than the 0.11 m thickness of the sample. Therefore, any improvement delivered by the UTR method in comparison with standard CT reconstruction methods like FBP, cannot be attributed to the incorporation of the Ewald sphere curvature, as in the first example above. However, there are two other factors that can explain the improved performance of the UTR algorithm. Firstly, UTR utilizes 3D gridding, which acts similarly to 3D linear interpolation in its handling of noise and spatial resolution in the data. On the other hand, the FBP method, and most other standard techniques for plane-wave CT, use 2D linear interpolation, which may be less efficient in filtering out the noise compared to the full 3D treatment in UTR. Secondly, our implementation of the UTR includes an internal noise filter in the Fourier space, which relies on the user-provided NSR parameter (see the description in the context of the first example above) to filter out the noise in the reconstructed Fourier coefficients $\mathbf{F}_3\delta(\mathbf{q})$. The UTR software suppresses the Fourier coefficients with magnitude below the threshold level of NSR $M/\sqrt{S(\mathbf{q})}$, where $M^2$ is the total number of pixels in each input image and the sampling matrix coefficients $S(\mathbf{q})$ were defined earlier in Section 2 after eq.(11). The NSR in the input images is multiplied by $M$ here in accordance with the well-known Parseval theorem of the Discrete Fourier Transform (DFT) [51] and the result is subsequently divided by $M^2$ after the inverse DFT. The division of the standard deviation of noise by the square root of the sampling coefficient is a consequence of the assumption about the noise being uncorrelated between different images.

## 4. Conclusions

We have developed a unified method for 3D reconstruction that can be applied in the same form in conventional (absorption-based), phase-contrast and diffraction tomography. The question about the relationship between these different CT modalities has been investigated previously [21], but to the best of our knowledge a "Fourier diffraction projection" theorem in the form of eq.(10) and the corresponding unified method for the reconstruction of the refractive index in the form of eq.(11) or similar, have not been reported previously. A clear advantage of this unified solution to the problem of CT reconstruction in the case of plane-wave illumination, compared to methods reported previously (see a discussion in the Introduction above), is the applicability of eq.(11) to standard CT scan data with a single input image per illumination direction. This makes the proposed method potentially highly practical and applicable to experimental conditions that are typically found in conventional and phase-contrast X-ray CT, including medical and biomedical CT, in high-resolution imaging with soft X-rays, in atomic-resolution TEM, including high-resolution cryo-EM, and possibly also in other popular imaging techniques using ultrasound, radio waves and other forms of radiation and matter waves for 3D imaging.

    We have also developed a fast 3D-gridding-based implementation of the UTR algorithm that can be applied to CT image data collected with plane-wave illumination in an arbitrary scanning geometry. Obviously, in order to obtain a unique reconstruction with a specified spatial resolution, the input data always needs to satisfy the Nyquist sampling conditions in general [5] and the Orlov-type conditions for the geometry of the set of incident illumination directions [35]. As long as these sampling conditions are satisfied, the UTR method is able to produce a unique and high-quality (in terms of the signal-to-noise and spatial resolution) 3D reconstruction from CT data collected along an arbitrary scanning trajectory. This "universality" of the proposed algorithm is based on the intrinsic calculation of the appropriate spatial filter that emerges as a result of the automatically detected sampling of the input data in the 3D reciprocal space. For example, in the simplest case of a conventional CT scan over 180 degrees around a fixed rotation axis, this in-built procedure intrinsically generates the usual ramp filter of the conventional CT [5]. We have presented two examples from different imaging domains which demonstrate that the developed implementation of the UTR method indeed performs as well as or better than the best specialized methods that have been reported and used in these domains to date.

    We have made our UTR software freely available [36] in the hope that it may be useful for researchers working in diverse fields of imaging physics.

Future work may be aimed at extension of the UTR method to CT geometries with non-planar illumination, such as e.g. cone-beam illumination that is utilized in the majority of industrial and medical CT scanners [5,20,52]. Given the rather simple and conceptually transparent structure of the reconstruction algorithm expressed by eq.(11), extensions of this method to fan-beam and cone-beam geometries appear straightforward. This can make the method applicable to CT imaging with conventional X-ray tubes and microfocus sources, where the divergence of the incident beam inside the object cannot be neglected [5,50,53].

**References**

1. R. N. Bracewell and A. C. Riddle, "Strip integration in radio astronomy," Aus. J. Phys. **9**, 198-217 (1956).
2. A. M. Cormack, "Representation of a function by its line integrals, with some radiological applications," J. Appl. Phys. **34**, 2722-2727 (1963).
3. R.A. Crowther, D.J. De Rosier and A. Klug, "The reconstruction of a three-dimensional structure from projections and its application to electron microscopy," Proc. R. Soc. London Ser. A **317**, 319-340 (1970).
4. G. N. Hounsfield, "Computerized transverse axial scanning tomography: Part I, description of the system," Br. J. Radiol. **46**, 1016-1022 (1973).
5. F. Natterer, *The Mathematics of Computerized Tomography* (SIAM, 2001).
6. J. Radon, "Uber die Bestimmung von Funktionen durch ihre Integralwerte langs gewisser Mannigfaltigkeiten," Berichte Sachsische Akademie der Wissenschaften, Leipzig, Math.—Phys. Kl. **69**, 262-267 (1917).
7. E. Wolf, "Three-dimensional structure determination of semi-transparent objects from holographic data," Opt. Commun. **1,** 153-156 (1969).
8. A. J. Devaney, "A filtered backpropagation algorithm for diffraction tomography," Ultrason. Imag. **4**, 336-350 (1982).
9. M. A. Anastasio and X. Pan, "Computationally efficient and statistically robust image reconstruction in three-dimensional diffraction tomography," J. Opt. Soc. Am. A **17**, 391–400 (2000).
10. G. Gbur, and E. Wolf, "Diffraction tomography without phase information," Opt. Lett. **27**, 1890–1892 (2002).
11. J. M. Cowley, *Diffraction Physics, 3rd ed.* (Elsevier, 1995).
12. J. P. J. Chen, K. E. Schmidt, J. C. H. Spence, and R. A. Kirian, "A new solution to the curved Ewald sphere problem for 3D image reconstruction in electron microscopy," Ultramicroscopy **224**, 113234 (2021).
13. K. H. Downing, and R. M. Glaeser, "Estimating the effect of finite depth of field in single-particle cryo-EM," Ultramicroscopy **184**, 94–99 (2018).
14. R. M. Glaeser, "How good can single-particle cryo-EM become? What remains before it approaches its physical limits?" Ann. Rev. Biophys. **48**, 45–61 (2019).
15. M. Bertilson, O. von Hofsten, U. Vogt, A. Holmberg, and H. M. Hertz, "High-resolution computed tomography with a compact soft x-ray microscope," Opt. Express **17**, 11057-11065 (2009).
16. A. Momose, "Demonstration of phase-contrast X-ray computed-tomography using an xray interferometer," Nucl. Instrum. Methods Phys. Res. A, 352, 622–628 (1995).
17. C. Raven, A. Snigirev, I. Snigireva, P. Spanne, A. Souvorov, and V. Kohn, " Phase-contrast microtomography with coherent high-energy synchrotron x rays," Appl. Phys. Lett. 69, 1826-1828 (1996).
18. P. Cloetens, M. Pateyron-Salome, J. Y. Buffiere, G. Peix, J. Baruchel, F. Peyrin, M. Schlenker, " Observation of microstructure and damage in materials by phase sensitive radiography and tomography," J. Appl. Phys. **81** 5878-5886 (1997).
19. A. V. Bronnikov, "Reconstruction formulas in phase-contrast tomography," Opt. Commun. **171**, 239-244 (1999).
20. S. W. Wilkins, Ya. I. Nesterets, T. E. Gureyev, S. C. Mayo, A. Pogany and A. W. Stevenson, "On the Evolution of X-Ray Phase-Contrast Imaging Methods and their Relative Merits," Phil. Trans. R. Soc. A **372**, 20130021 (2014).
21. M. A. Anastasio and D. Shi, "On the relationship between intensity diffraction tomography and phase-contrast tomography," Proc. SPIE **5535**, 361-368 (2004).
22. D. M. Paganin, *Coherent X-ray Optics* (Clarendon Press, 2006).
23. A. Momose and J. Fukuda, "Phase-contrast radiographs of nonstained rat cerebellar specimen," Med. Phys. **22**, 375-379 (1995).
24. T. E. Gureyev, H. M. Quiney and L. J. Allen, "A method for virtual optical sectioning and tomography utilizing shallow depth of field," J. Opt. Soc. Am. A **39**, 936-947 (2022).
25. T. E. Gureyev, Y. I. Nesterets, D. M. Paganin, A. Pogany, and S. W. Wilkins, "Linear algorithms for phase retrieval in the Fresnel region. 2. Partially coherent illumination," Opt. Commun., **259**, 569-580 (2006).
26. D. Paganin, S. C. Mayo, T. E. Gureyev, P. R. Miller, and S. W. Wilkins, "Simultaneous phase and amplitude extraction from a single defocused image of a homogeneous object," J. Microsc. **206**, 33–40 (2002).
27. T. E. Gureyev, Ya. I. Nesterets, and D. M. Paganin, (2015) Monomorphous decomposition method and its application for phase retrieval and phase-contrast tomography, Phys. Rev. A **92**, 053860 (2015).
28. T. E. Gureyev, T. J. Davis, A. Pogany, S. C. Mayo, and S. W, Wilkins, "Optical phase retrieval by use of first Born- and Rytov-type approximations," Applied Optics **43**, 2418-2430 (2004).

29. A. Pogany, D. Gao, and S. W. Wilkins, "Contrast and resolution in imaging with a microfocus x-ray source," Rev. Sci. Instrum. **68**, 2774-2782 (1997).
30. T. E, Gureyev, H. M. Quiney, A. Kozlov, and L. J. Allen, "Relative roles of multiple scattering and Fresnel diffraction in the imaging of small molecules using electrons," Ultramicroscopy **218**, 113094 (2020).
31. D. Paganin, T. E. Gureyev, S. C. Mayo, A. W. Stevenson, Y. I. Nesterets, and S. W. Wilkins, "X-ray omni microscopy," J. Micros. **214**, 315-327 (2004).
32. D. Shi and M. A. Anastasio, "Intensity diffraction tomography with fixed detector plane," Opt. Engineer. **46**, 107003 (2007)..
33. M. Defrise, R. Clack and D. Townsend, "Solution to the three-dimensional image reconstruction problem from two-dimensional parallel projections," J. Opt. Soc. Am. A **10**, 869-877 (1993).
34. P. A. Penczek, R. Renka and H. Schomberg, "Gridding-based direct Fourier inversion of the three-dimensional ray transform," J. Opt. Soc. Am. A **21**, 499-509 (2004).
35. S. S. Orlov, "Theory of three-dimensional reconstruction. 1. Conditions of a complete set of projections," Sov. Phys. Crystallogr. **20**, 312–314 (1975).
36. T. E. Gureyev, "Unified Tomographic Reconstruction algorithm," UTR_public, Github, (2022), https://github.com/timg021/UTR_public.
37. H. G. Brown, Z. Chen, M. Weyland, C. Ophus, J. Ciston, L. J. Allen, and S. D. Findlay, "Structure retrieval at atomic resolution in the presence of multiple scattering of the electron probe," Phys. Rev. Lett. **121**, 266102 (2018).
38. D. Ren, C. Ophus, M. Chen, and L. Waller, "A multiple scattering algorithm for three dimensional phase contrast atomic electron tomography," Ultramicroscopy **208**, 112860 (2019).
39. D. Thompson, Ya. I. Nesterets, K. M. Pavlov and T. E. Gureyev, "Fast three-dimensional phase retrieval in propagation-based x-ray tomography," J. Synchrot. Radiat. **26**, 1-14 (2019).
40. D. Thompson, Ya. I. Nesterets, K. M. Pavlov and T. E. Gureyev, "Three-dimensional contrast transfer functions in propagation-based tomography," Eprint arXiv: 2206.02688 (2022).
41. J. Zhou, Yo. Yang, Ya. Yang, D. S. Kim, A. Yuan, X. Tian, C. Ophus, F. Sun, A. K. Schmid, M. Nathanson, H. Heinz, Q. An, H. Zeng, P. Ercius, and J. Miao, "Observing Nucleation in 4D - particle1 (measurement 1)," Materials Data Bank (2019), https://www.materialsdatabank.org/dataset/FePt00002.
42. T. E. Gureyev, D. M. Paganin, H. G. Brown, H. M. Quiney, and L. J. Allen, "A method for high-resolution three-dimensional reconstruction with Ewald sphere curvature correction from transmission electron images," accepted for publication in Microscopy and Microanalysis (2022).
43. T. E. Gureyev, "Sparse structures TEM," Github (2022), https://github.com/timg021/SparseStructuresTEM2.
44. K. Momma, and F. Izumi, "An integrated three-dimensional visualization system VESTA using wxWidgets," IUCr Newslett. **7**, 106-119 (2006).
45. E. J. Kirkland, Advanced Computing in Electron Microscopy, 2nd ed. (Springer, 2010).
46. E. J. Kirkland, "Temsim", Github (2022), https://github.com/jhgorse/kirkland/tree/master/temsim.
47. E. J. Kirkland, "Computem: Transmission Electron Microscope Image Simulation", SourceForge (2022), https://sourceforge.net/projects/computem.
48. T. E. Gureyev, Ya. I. Nesterets, P. M. Baran, S. T. Taba, S. C. Mayo, D. Thompson, B/ Arhatari, A. Mihocic, B. Abbey, D. Lockie, J. Fox, B. Kumar, Z. Prodanovic, D. Hausermann, A. Maksimenko, C. Hall, A. G. Peele, M. Dimmock, K. M. Pavlov, M. Cholewa, S. Lewis, G. Tromba, H. M. Quiney, and P. C. Brennan, "Propagation-based X-ray phase-contrast tomography of mastectomy samples," Med. Phys. **46**, 5478-5487 (2019).
49. F. A. Stewart, A. V. Akleyev, M. Hauer-Jensen, J. H. Hendry, N. J. Kleiman, T. J. MacVittie, B. M. Aleman, A. B. Edgar, K. Mabuchi, C. R. Muirhead, R. E. Shore, W. H. Wallace, "CRP Publication 118: ICRP Statement on Tissue Reactions and Early and Late Effects of Radiation in Normal Tissues and Organs – Threshold Doses for Tissue Reactions in a Radiation Protection Context," Annals of the ICRP **41**, 1-322 (2012).
50. T. E. Gureyev, Ya. I. Nesterets, D. Thompson, S. W. Wilkins, A. W. Stevenson, A. Sakellariou, J. Taylor, and D. Ternovski, "Toolbox for advanced X-ray image processing," Proc. SPIE **8141**, 81410B-1 - 81410B-14 (2011).
51. W. H. Press, S. A. Teukolsky, W. T. Vetterling and B. R. Flannery, *Numerical Recipes in C, The Art of Scientific Computing. 2nd Edition* (Cambridge University Press, 1992).
52. J. Hsieh and T. Flohr, "Computed tomography recent history and future perspectives," J. Med. Imag. **8**, 052109 (2021).
53. L.A. Feldkamp, L.C. Davis, and J.W. Kress, "Practical cone-beam algorithm," J. Opt. Soc. Am. A **1**, 612-619 (1984).